\documentclass[prl,preprint,superscriptaddress,showpacs,endfloats*]{revtex4-1}

\usepackage{amsmath}
\usepackage{amssymb}
\usepackage{graphicx}

\newcommand{\T}{${\mathcal T}\,$}
\newcommand{\Ti}{${\mathcal T}$}

\begin{document}

\title{Exceptional Points in a Microwave Billiard with \\ 
       Time-Reversal Invariance Violation}

\author{B.~Dietz}
%\email{dietz@ikp.tu-darmstadt.de}
\affiliation{Institut f{\"u}r Kernphysik, Technische Universit{\"a}t
Darmstadt, D-64289 Darmstadt, Germany}

\author{H.~L.~Harney}
%\email{hanns-ludwig.harney@mpi-hd.mpg.de}
\affiliation{Max-Planck-Institut f{\"u}r Kernphysik, D-69029 Heidelberg,
Germany}

\author{O.~N.~Kirillov}
\affiliation{FWSH, Helmholtz-Zentrum Dresden-Rossendorf,
             P.O. Box 510119, D-01314 Dresden, Germany}

\author{M.~Miski-Oglu}
%\email{maksim@ikp.tu-darmstadt.de}
\affiliation{Institut f{\"u}r Kernphysik, Technische Universit{\"a}t
Darmstadt, D-64289 Darmstadt, Germany}

\author{A.~Richter}
\email{richter@ikp.tu-darmstadt.de}
\affiliation{Institut f{\"u}r Kernphysik, Technische Universit{\"a}t
Darmstadt, D-64289 Darmstadt, Germany}
\affiliation{$\rm ECT^*$, Villa Tambosi, I-38123 Villazzano (Trento), Italy}

\author{F.~Sch{\"a}fer}
%\email{schaefer@ikp.tu-darmstadt.de}
\affiliation{Institut f{\"u}r Kernphysik, Technische Universit{\"a}t
Darmstadt, D-64289 Darmstadt, Germany}
\affiliation{LENS, University of Florence, I-50019 Sesto-Fiorentino (Firenze),
Italy}

\date{\today}

\begin{abstract}
        We report on the experimental study of an exceptional point (EP) in a 
        dissipative microwave billiard with induced time-reversal invariance (\Ti) violation. 
	The associated two-state Hamiltonian is non-Hermitian and non-symmetric. 
        It is determined experimentally on a narrow grid in a parameter plane 
        around the EP. At the EP the size of \T violation 
	is given by the relative phase of the eigenvector components. The eigenvectors
	are adiabatically transported around the EP, whereupon they gather  
	geometric phases and in addition geometric amplitudes
        different from unity. 
\end{abstract}


\maketitle

We present experimental studies of two nearly degenerate eigenmodes 
in a dissipative microwave cavity with induced \Ti
violation. Due to the dissipative nature of the system the associated Hamiltonian is 
not Hermitian~\cite{GW88,MF80,HH04,R08} and thus 
may possess an exceptional point (EP), where two or more eigenvalues and also the associated 
eigenvectors coalesce. In contrast, at a degeneracy of a Hermitian Hamiltonian, a so-called 
diabolical point (DP), the eigenvectors are linearly independent 
\cite{B84,BD03}. 
The occurrence of exceptional points \cite{Ka66,MF80} in the spectrum of a dissipative 
system 
was studied in quantum physics \cite{La95} as well as in classical physics
\cite{EPTh09}. It has been demonstrated, that this
is not only a mathematical but also a physical phenomenon. 
The first experimental evidence of EPs came
from flat microwave cavities \cite{Phillip,D01,Dembo2003,Metz}, which are analogues of quantum 
billiards \cite{QB}.
Subsequently EPs were observed in coupled electronic circuits
\cite{SHS04} and recently in chaotic microcavities and atom-cavity quantum composites 
\cite{Lee2009}. The present contribution is the first experimental 
study of an EP under \T violation. It is induced via magnetization of a 
ferrite 
inside the cavity by an external field $\rm B$~\cite{Schaefer2007}.  
\T violation caused by B is commonly distinguished from dissipation \cite{Haake}. For a 
non-dissipative system with broken \T invariance (${\rm B}\ne 0$) the 
Hamiltonian is Hermitian. For a dissipative system the energy is not 
conserved such that for ${\rm B}=0$ it is described by a complex symmetric 
Hamiltonian. This case is referred to as the \Ti-invariant one \cite{Haake}. 

To realize a coalescence of a doublet of eigenmodes in the experiment two parameters are varied. 
Since the doublets considered are
well separated from neighboring resonances, the effective Hamiltonian is two-dimensional. 
Its four complex elements are determined on 
a narrow grid in the parameter plane, thus yielding an unprecedented
set of data. This allows (i) to quantify the size of \T violation, (ii) to measure 
to a high precision the geometric phase \cite{B84,BD03} and the geometric amplitude 
\cite{GW88,MKS05,MM08} 
that the eigenvectors gather when encircling an EP.  
An earlier rule \cite{D01} on the encircling is generalized to 
the case of \T violation \cite{GW88}. 

\begin{figure}[ht]
	\centering
	\includegraphics[width=6cm]{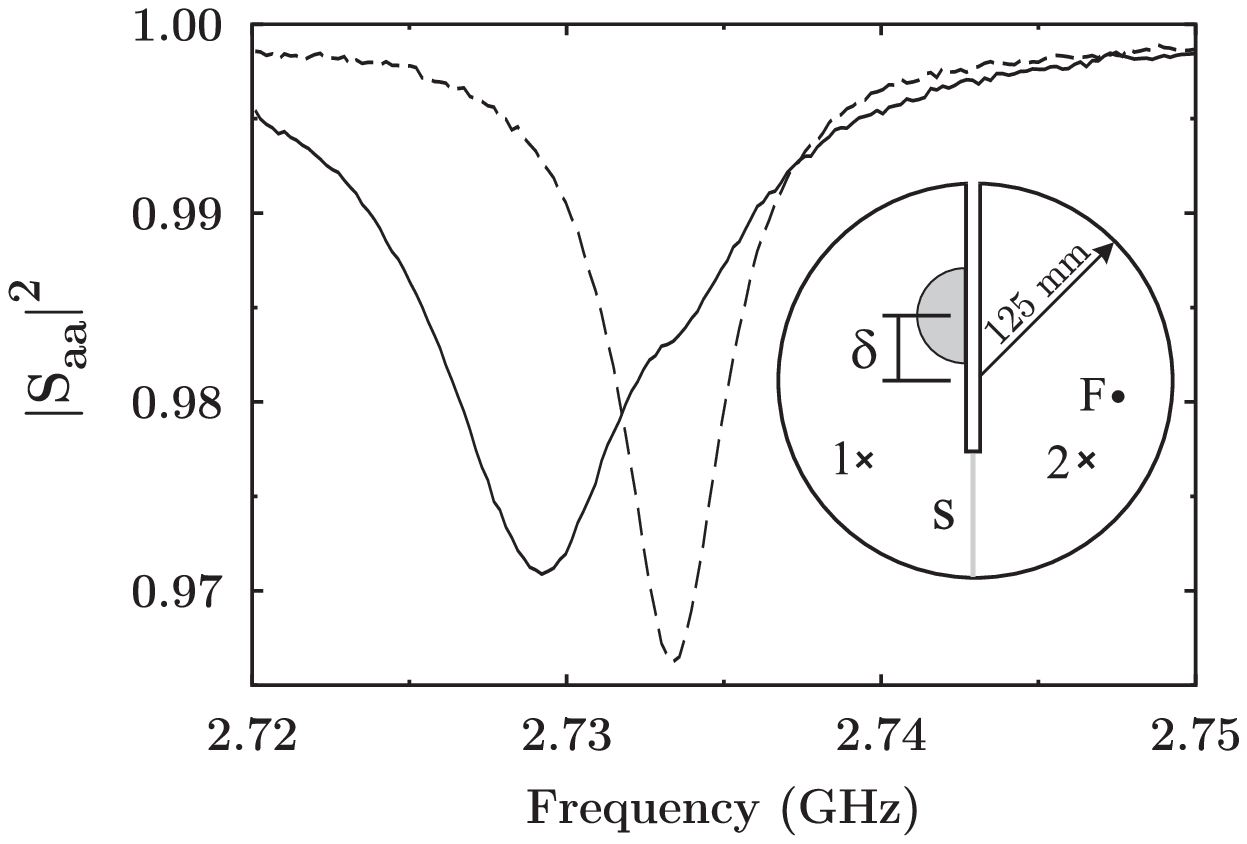}
	\caption{Reflection spectra measured at antenna $1$ (solid) 
                 and $2$ (dashed) for $s = 1.66~{\rm mm}$, 
                 $\delta = 41.50~{\rm mm}$ and ${\rm B}=53~{\rm mT}$. 
                 Each antenna couples to predominantly one eigenmode for 
                 that choice of parameters. We observe a single resonance at 
                 antenna 2, a much broader with a shoulder at the peak position                 
                 of the former at antenna 1. Thus one eigenmode is localized 
                 in the right part of the cavity, whereas the second 
                 one penetrates from the left into the right part of the 
                 cavity. 
                 Inset: top view (to scale) of the microwave cavity. In each 
                 half of the resonator an antenna, $1$ and $2$, is positioned. 
                 A semicircular Teflon disk (gray) is positioned at a distance
                 $\delta$ from the center, $s$ refers to the height of 
                 the opening (gray bar) between both cavity parts. 
                 A ferrite is located at $F$.}
	\label{fig:1}
\end{figure}

The experimental setup is similar to that used in \cite{D01,Dembo2003}. 
The resonator is constructed from three 5~mm thick copper plates, which are sandwiched 
on top of each other. The center plate has a hole with the shape of two half circles of 
250~mm in diameter, which are separated by a 10~mm bar of copper except for an opening
of 80~mm length (see the inset of Fig. \ref{fig:1}).
The opening allows a coupling of the electric field modes excited in each half circle,
which is varied with a movable gate of copper of length 80~mm and 
width 3~mm inserted through a small opening in the top plate and operated by 
a micrometer stepper motor. The bottom of this gate is tilted to allow a precise 
setting of small couplings. When closing the gate it eventually plunges 
into a notch in the bottom plate. Its lifting $s$  
defines one parameter with the range 
$0~{\rm mm\ (no\ coupling)} \leq s \leq 9~{\rm mm\ (full\ coupling)}$.
The left half of the circular cavity in Fig.~\ref{fig:1} contains a movable semicircular 
Teflon piece with diameter 60~mm and height 5~mm connected to the outside by a thin
snell operated by another stepper motor. Its displacement with respect 
to the center of the cavity defines the other parameter $\delta$. 

Two pointlike dipole antennas intrude 
into each part of the cavity. A vectorial network analyzer (VNA) Agilent PNA 5230A couples microwave 
power into the cavity through one antenna $a$ and measures amplitude and phase of the signal received at the same
(reflection measurement) or the other (transmission measurement) antenna $b$ 
relative to the input signal. In this way the 
complex element $S_{ba}$ of the scattering matrix $S$ is determined. 
To induce \T violation, a ferrite with the shape of a cylinder of
diameter 4~mm and height 5~mm~\cite{Schaefer2007} is placed in the
right part of the cavity and magnetized from the outside by two 
permanent magnets. They are mounted to a screw thread mechanism above and below the resonator
and magnetic field strengths $0~{\rm mT} \leq {\rm B} \leq 90~{\rm mT}$ are obtained by varying 
their distance. To automatically 
scan the parameter space spanned by $(s, \delta)$, the two stepper motors and 
the VNA are controlled by a PC. The four $S$-matrix elements 
$S_{ba}(f)$, $\{a,b\} \in \{1,2\}$ are measured in the parameter plane  
with the resolution $\Delta s = \Delta \delta = 0.01~{\rm mm}$ and a 
frequency step $\Delta f =10~{\rm kHz}$. 
The frequency range of 40~MHz is determined by the spread of the resonance 
doublet. 
Lack of reciprocity, i.e. $S_{12}\ne S_{21}$, is the signature for \T violation~\cite{Schaefer2007}.
Figure~\ref{fig:1} shows two typical reflection spectra, i.e. 
$\vert S_{aa}\vert^2$ with $a=1,2$. Additional measurements include
neighboring resonances, which are situated about $250~{\rm MHz}$ away from the region of
interest, in order to account for their residual influence. In the considered frequency range the 
electric field vector is 
perpendicular to the top and bottom plates. Then, the Helmholtz equation
is identical to the Schr\"odinger equation 
for a quantum billiard of corresponding shape \cite{QB,Richter}. Consequently, the 
results provide insight into the associated quantum problem.

\begin{figure}[ht]
	\centering
	\includegraphics[width=8cm,height=3.5cm]{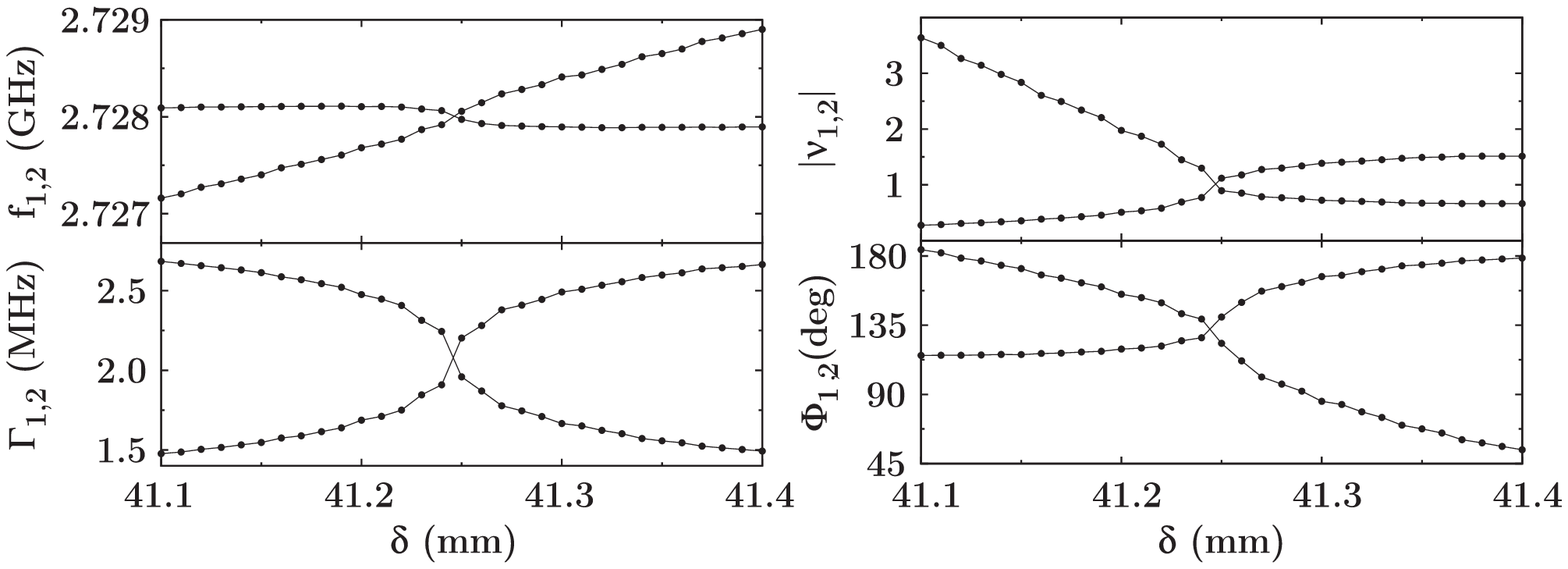}
	\caption{Left panels: Real and imaginary parts of the eigenvalues, $f_{1,2}$ and $\Gamma_{1,2}$, 
        as function of $\delta$ at $s=s_{EP}=1.66~{\rm mm}$ and 
        ${\rm B} = 53~{\rm mT}$. Around $\delta=\delta_{EP}=41.25~{\rm mm}$ they are closest. 
	Right panels: Modulus and phase of the ratios $\nu_{1,2}=
	|\nu_{1,2}|e^{i\Phi_{1,2}}$ of the 
	components of the associated eigenvectors.
        They are also closest at $\delta\simeq 41.25$~mm. For $\delta\leq\delta_{EP}$ the upper curves 
        correspond to,
        respectively $f_1,\, \delta_1,\, |\nu_2|,\, \Phi_2$.}
	\label{fig:2}
\end{figure}
In previous experiments \cite{D01,Dembo2003} an EP was located by 
determining for each parameter setting the real and imaginary parts of the eigenvalues as 
the frequencies $f_{1,2}$ and the widths $\Gamma_{1,2}$ of the resonances from a fit of a 
Breit-Wigner function. This procedure, however, fails at the EP because there 
the line shape is not  
a first order pole of the $S$ matrix but rather the sum 
of a first and a second order pole \cite{Metz,R08}. Therefore, the coalescence of 
the eigenvectors should be incorporated in the search for the EP 
\cite{Dembo2003,Metz}. For this we determine the 
two-state Hamiltonian $H$ and 
its eigenvalues and eigenvectors explicitely for every setting of the parameters $s,\delta ,{\rm B}$ 
from the measured $S$-matrix elements via the method presented 
in \cite{Schaefer2007}. There, we showed that for a resonator with two pointlike antennas 
a resonance doublet is well described by the two-channel $S$ matrix 
$S(f)=\boldsymbol 1-2\pi iW^{\dagger}(f\, \boldsymbol 1-H)^{-1}W\,$.
The matrix $W=(W_{\mu a})$ couples the
resonant states $\mu =1,2$ to the antennas $a=1,2$. It is real
since that coupling conserves \T. The Hamiltonian $H$ comprises dissipation 
in the walls of the resonator and the ferrite and 
\T violation and thus is neither Hermitian nor symmetric, which implies 
$S_{12}\ne S_{21}$. Its general form is given as
\begin{equation}
H=\left(\begin{array}{cc}
         e_1\,          &H^S_{12}-iH^A_{12}\\
         H^S_{12}+iH^A_{12}\,         &e_2
        \end{array}
  \right)\, .
                   \label{eq:2}
\end{equation}
The quantities $e_1\pm e_2, H^S_{12}$ and $H^A_{12}$ are complex expansion coefficients with respect 
to the unit and the Pauli matrices. The ansatz for $S$ was tested thoroughly by fitting it 
to $S$-matrix elements measured with and without magnetization of the ferrite.
 In the latter case the antisymmetric part 
$H^A_{12}$ vanishes and $H$ coincides with that used in \cite{D01,Dembo2003}.
Fitting the $S$ matrix to the 
four measured excitation functions $S_{ba}(f)$ yields the matrices $H$ and $W$ up to common real 
orthogonal basis transformations. We define the basis 
such that $(H^S_{12}+iH^A_{12})/(H^S_{12}-iH^A_{12})=\exp (2i\tau)$ is a phase factor.
Thus, \T violation is expressed by a real phase $\tau$. This is usual practice 
in physics, e.g., for nuclear reactions \cite{Driller}, and for weak and 
electromagnetic decay \cite{Ri75}. 

\begin{figure}[ht]
	\centering
	\includegraphics[width=6cm]{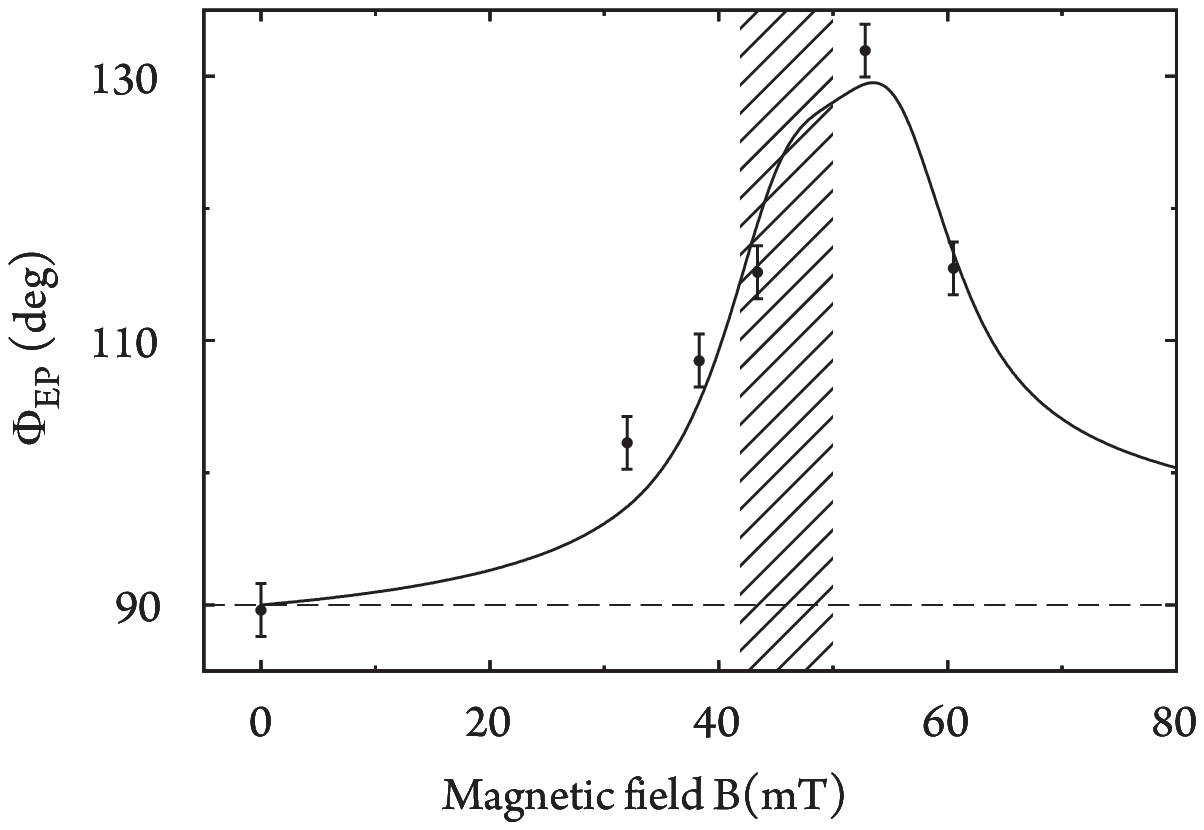}
	\caption{Phase of the ratio $\nu_{EP}$ of the eigenvector components 
        at the EP as a function of the magnetization of the ferrite. 
        For the \Ti-conserving case the known result~\cite{Dembo2003} of $90^\circ$ 
	(dashed horizontal line) is recovered. The model for the \Ti-violating matrix
	element $iH^A_{12}$ in terms of the ferromagnetic resonance yields the solid line which
        agreeably describes the data. The vertical bar indicates the range of ${\rm B}$ 
        where the ferromagnetic resonance is expected.}
	\label{fig:3}
\end{figure}

The eigenvalues of $H$ in Eq.~(\ref{eq:2}) coalesce to an EP when 
${H^S_{12}}^2+{H^A_{12}}^2+(e_1-e_2)^2/4=0$ but not all three terms equal to 
zero. Figure~\ref{fig:2} shows one example for a search of the EP. 
Keeping $s,{\rm B}$ fixed at $s=1.66$~mm and ${\rm B}=53$~mT, the real and imaginary parts of the
eigenvalues, $f_j$ and $\Gamma_j$ are shown as functions of $\delta$. At $\delta =41.25$~mm the 
encounter of the eigenvalues is closest. To determine the parameter values for the 
coalescence of the eigenvectors $|r_k\rangle,\,\, k=1,2,$ their components $r_{k1},r_{k2}$  
and the ratios $\nu_k=r_{k1}/r_{k2}=|\nu_k|e^{i\Phi_k}$ are determined \cite{BD03}.  
The right part of Fig.~\ref{fig:2} demonstrates that the encounter of the eigenvectors is also closest 
at $\delta =41.25$~mm. Indeed, as will be evidenced in further tests below, both the eigenvalues
and the eigenvectors cross at these parameter values. The EP is located at 
$(s_{EP},\delta_{EP})=(1.66\pm 0.01,41.25\pm 0.01)$~mm. At the EP the only 
eigenvector of $H$ is 
\cite{HH04}
\begin{equation}
|r_{\rm EP}\rangle=\left(\begin{array}{c}
                           i\, e^{i\tau}\\
                           1
                        \end{array}
                  \right)\, .
                           \label{eq:4}
\end{equation}
The ratio of its components is a phase factor. For \Ti-conserving systems  
the phase is $\Phi_{\rm EP}=\pi/2$ \cite{Dembo2003} as confirmed by Fig.~\ref{fig:3} at ${\rm B}=0$. 
With \T violation the phase equals $\Phi_{\rm EP}=\tau+\pi/2$ and
thus provides a measure for its size $\tau$. Figure~\ref{fig:3} shows that with increasing 
${\rm B}$ the parameter $\tau$
goes through a maximum. As in Ref.~\cite{Schaefer2007} we identify this with the 
ferromagnetic resonance resulting from the coupling of the 
rf~magnetic field to the spins in the ferrite. The \Ti-violating matrix element $iH^A_{12}$ 
has been expressed by a resonance formula which yields the solid curve for $\Phi_{\rm EP}({\rm B}$). 

\begin{figure}[ht]
    \setlength{\unitlength}{1mm}
	\centering
    \includegraphics[width=8cm]{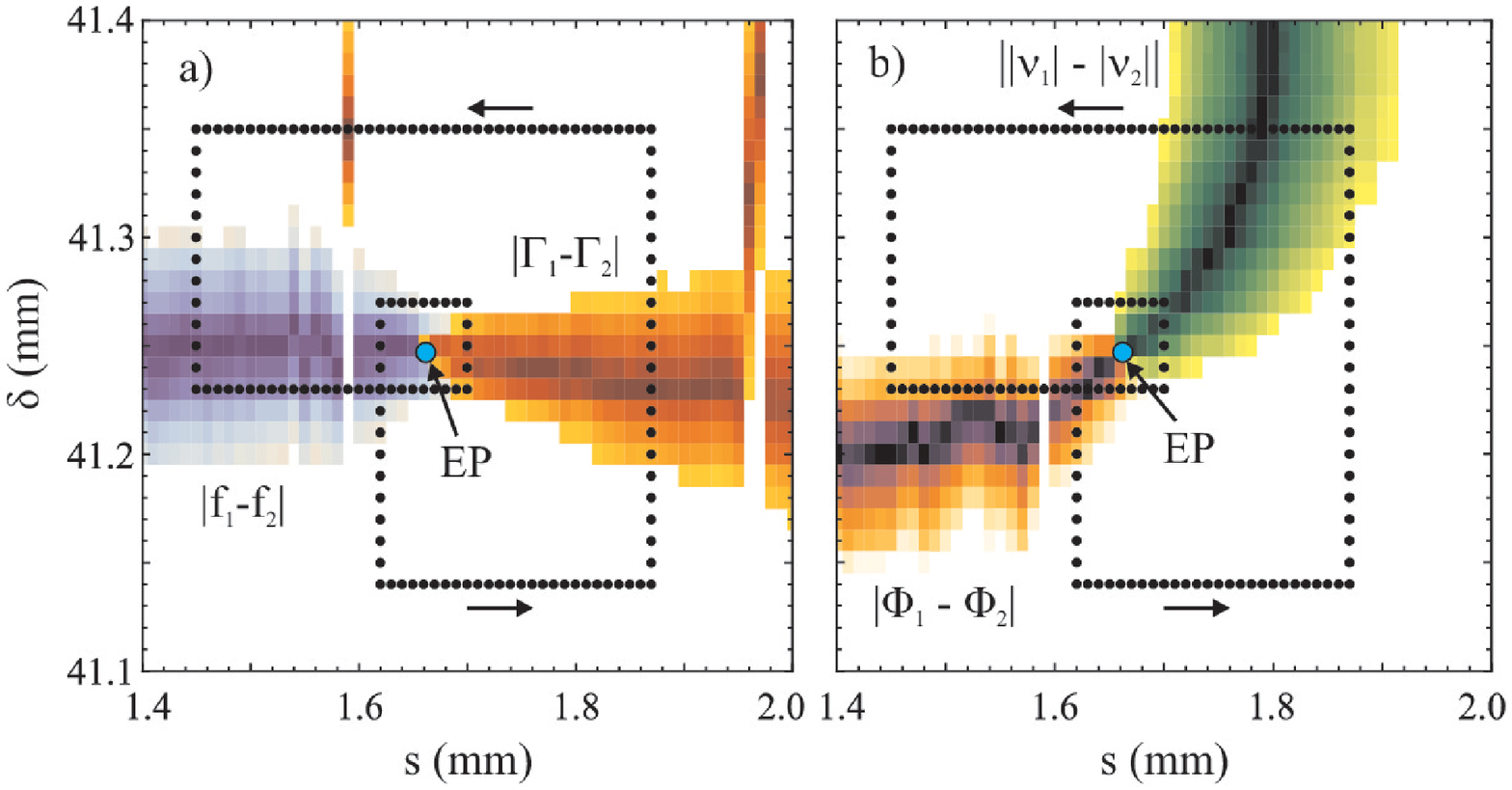}
	\caption{Differences of the complex eigenvalues $f_{1,2}+i\Gamma_{1,2}$ (a) and of the
	ratios $\nu_{1,2}=\vert\nu_{1,2}\vert e^{i\Phi_{1,2}}$  of the 
	eigenvector components (b) in an area of the parameter plane $(s,\delta)$ around the EP.
	The darker the colour the smaller is the respective difference.
        That of $f_{1,2}$ (blue) and of $\Phi_{1,2}$ (orange) are small only 
	to the left, those of $\Gamma_{1,2}$ (orange) and of $|\nu_{1,2}|$ (green) to the right 
	of the EP. The black dots in both panels indicate the chosen contour for the encircling 
	of the EP (see text).}
	\label{fig:4}
\end{figure}

Panel~(a) of Fig.~\ref{fig:4} shows in a neighborhood of the EP at 
${\rm B}=53$ mT the differences of the complex eigenvalues,
$f_{1,2}$ (blue, left to EP) and $\Gamma_{1,2}$ (orange, right to EP), panel~(b) those of 
the phases $\Phi_{1,2}$ (orange, left to EP) and of the moduli 
$\vert\nu_{1,2}\vert$ (green, right to EP). 
The darker the colour the smaller is the respective difference. The darkest 
colour visualizes the curve (branch cut) along which it vanishes. The left and 
right panels of Fig.~\ref{fig:2} are cuts through the respective panels of 
Fig.~\ref{fig:4} at $s=1.66$~mm. We observe that $|f_1-f_2|$ and 
$|\Phi_1-\Phi_2|$ are small and thus visible only to the left of the EP, 
whereas $|\Gamma_1-\Gamma_2|$ and 
$||\nu_1|-|\nu_2||$ are visible only to its right. Thus the branch cuts 
all  extend from one common point into opposite directions. This proves that 
this point is an EP \cite{BD03,MF80,Lee2009}. 

For systems with \T violation the geometric phase $\gamma$ gathered by the
eigenvectors around an EP is predicted to be 
complex yielding a geometric amplitude $e^{-{\rm Im}\gamma}\ne 1$ \cite{GW88}.
To check this we choose contours around the EP for the six values of the 
magnetic field ${\rm B}$ 
considered in Fig.~\ref{fig:3}. One example is shown in Fig.~\ref{fig:4}. 
As proposed in \cite{MM08} it consists of two different loops. 
The path is parametrized by a real variable $t$ with initial value $t=0$. 
In \cite{Dubbers} and \cite{D01} the geometric phase gathered 
around a DP, respectively an EP, was obtained for just a few parameter settings, because the procedure -- the measurement of the electric 
field intensity distribution -- is very time consuming. We 
now have the possibility to determine the left and right eigenvectors, 
$\langle l_j(t)|$ and $|r_j(t)\rangle,\, j=1,2$ of $H$ in Eq.~(\ref{eq:2}) on a much 
narrower grid of the parameter plane. At each point $t$ of 
the contour they are biorthonormalized, such that 
$\langle l_j(t)|r_j(t)\rangle =1$. 
Defining in analogy to the \Ti-conserving case \cite{D01}
$\mathcal{B}=\frac{(e_1-e_2)}{2}/ \sqrt{{H^S_{12}}^2+{H^A_{12}}^2},\,  
\tan\theta =\sqrt{1+\mathcal{B}^2} -\mathcal{B}$ yields for the right 
eigenvectors~\cite{GW88}
\begin{equation}
|r_1(t)\rangle=\left(\begin{array}{c}
                           e^{-i\tau/2}\cos\theta\\
                           e^{i\tau/2}\sin\theta 
                        \end{array}
                  \right),\,
|r_2(t)\rangle=\left(\begin{array}{c}
                           -e^{-i\tau/2}\sin\theta\\
                           e^{i\tau/2}\cos\theta
                        \end{array}
                  \right)\, .
                           \label{eq:6}
\end{equation}
As in \cite{D01} the EPs are located at $1+\mathcal{B}^2=0$ and 
encircling the EP once changes $\theta$ to $\theta\pm\pi /2$. 
The \Ti-violating parameter $\tau$ is not constant along the contour, even 
though the 
magnetic field is fixed. In fact, it varies with the opening $s$ between 
both parts of the 
resonator, because the ferrite is positioned in one of them, see 
Fig. \ref{fig:1}. The position of the EP does not depend on the value of $\tau$. 
Thus, the space curve $(s(t),\delta(t),\tau(t))$ winds around 
the line $(s_{EP},\delta_{EP},\tau(t))$, see Ref.~\cite{MKS05}. 
Since $\tau$ has no singular 
points in the considered parameter plane it returns to its 
initial value after each encircling of the EP. As a consequence, the 
eigenvectors $|r_{1,2}(t)\rangle$ follow the same 
transformation scheme as in the \Ti-conserving case. 
After completing the first loop at $t=t_1$ we have 
$|r_1(t_1)\rangle=|r_2(0)\rangle$, $|r_2(t_1)\rangle=-|r_1(0)\rangle$ and after the second 
one at $t=t_2$ we find $|r_{1,2}(t_2)\rangle=-|r_{1,2}(0)\rangle$. 

\begin{figure}[ht]
	\centering
\includegraphics[width=6.5cm]{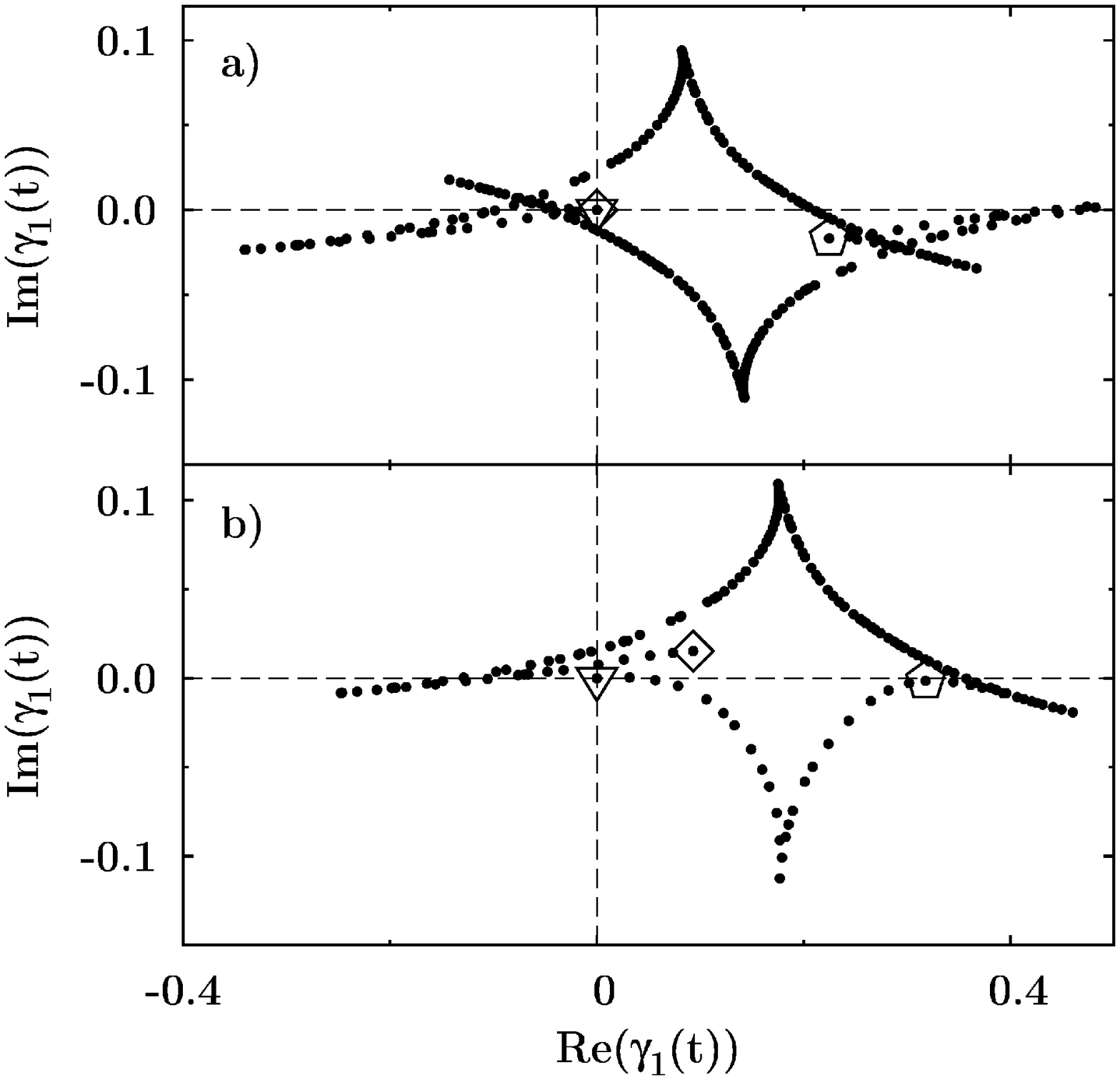}
	\caption{The complex phase $\gamma_1(t)$ 
        for ${\rm B} = 53~{\rm mT}$ when encircling the EP twice along the outer loop 
	(upper panel), and along the inner and outer loops (lower panel) of the contour shown in Fig.~\ref{fig:4}. 
	The triangle marks the start point, the pentagon the point after encircling 
	the EP once ($t=t_1$), the diamond that after a second encircling ($t=t_2$). There, 
	$\gamma_1(t_2)$ equals $(3.123-i3.474)\cdot 10^{-7}$, i.e. it is close to zero in the upper case, 
	$0.0931+i0.0152$ in the lower one.} 
	\label{fig:5}
\end{figure}

The biorthonormality defines the eigenvectors $\langle l_j|$ and $|r_j\rangle$ up to 
a geometric factor, so that $\langle L_j(t)|=\langle l_j(t)|\, e^{-i\gamma_j(t)}$ and $|R_j(t)\rangle=|r_j(t)\rangle\, e^{i\gamma_j(t)}$.
The geometric phases $\gamma_j(t)$ are fixed by the condition of parallel transport 
\cite{B84,GW88,FN}, $\left\langle L_j(t)|\frac{\rm d}{{\rm d}t}R_j(t)\right\rangle=0$.
This yields
$\frac{{\rm d}\gamma_1(t)}{{\rm d}t}=\frac{1}{2}\cos 2\theta(t)\frac{{\rm d}\tau(t)}{{\rm d}t}=
-\frac{{\rm d}\gamma_2(t)}{{\rm d}t}$. 
The initial value of the phases is chosen as $\gamma_j(0)=0$, such that $\gamma_1(t)=-\gamma_2(t)$. For the \Ti-conserving case we find $\gamma_{1,2}(t)\equiv 0$. 
The phase $\gamma_1(t)$ is   
determined successively for increasing $t$ from the product of $\langle l_1(t)|$ and $\left\vert\frac{\rm d}{{\rm d}t}r_1(t)\right\rangle$. 
In the upper panel of Fig.~\ref{fig:5} is plotted the phase $\gamma_1 (t)$ accumulated by $|R_1(t)\rangle$ when encircling the EP twice along the outer loop of the contour in Fig.~\ref{fig:4}. 
The lower panel shows $\gamma_1(t)$ for the contour in Fig.~\ref{fig:4}. 
The start point is the intersection of the loops. The orientation is chosen
such that the EP is always to the left.  
The cusps occur where $\frac{{\rm d}\tau}{{\rm d}t}=0$. In each panel the triangle marks the start point, the pentagon the point $t_1$, 
where the EP is encircled once, the diamond marks the point $t_2$ of completion of the second 
encircling. One sees that in both examples 
$\gamma_1(t_1)\ne 0$ and that, as predicted in Ref.~\cite{GW88},  
the geometric phase $\gamma_1(t_1)$ is not real.
If the EP is encircled twice along the same loop, we obtain 
$\gamma_1(t_2)=0$~\cite{MKS05,MM08} within $10^{-7}$. Thus we measured the $\gamma$'s up to an accuracy 
manifestly 
better than $10^{-2}$. Accordingly, from the value of $\gamma_1(t_2)$ given in the caption of Fig.~\ref{fig:5} 
we may 
conclude that, when the loops are different $\gamma_1(t)$ does not return to 
its initial value \cite{MM08}. When encircling this double loop again and
again we observe a drift of $\gamma_{1,2}$ in the complex plane away from the 
origin. Thus, one geometric amplitude increases and the other one decreases. 
This provides the experimental proof 
for the existence of a reversible geometric pumping process, as predicted in~\cite{GW88}. 
The 
dependence on the choice of the path signifies that the complex phases $\gamma_{1,2}(t)$ are geometric 
and not topological. In order to obtain the complete 
phase gathered by $|R_1(t)\rangle$ during the encircling of the EP one has to add the topological 
phase accumulated by $|r_1(t)\rangle$. 

In summary, when encircling the EP we obtain geometric factors different
from unity and the transformation scheme Eq.~(\ref{eq:6}) for $\vert\vec r_{1,2}(t)\rangle$.
These results provide an unambiguous proof that the EP lies inside both loops 
of the contour shown in Fig.~\ref{fig:4}. 
Their precision is illustrated by the dense sequence of points 
in Fig.~\ref{fig:5}. The accuracy of the measurements at and around the EP 
allows the determination of the size of \T violation and of the geometric factor along 
arbitrary contours in the parameter plane. Predictions on geometric amplitudes 
and phases could be confirmed. 

\begin{acknowledgments}
We thank G. Ripka and H.~A.~Weidenm\"uller for illuminating discussions.
This work was supported by the DFG within SFB~634 and by the Alexander-von-Humboldt Foundation.
\end{acknowledgments}


\begin{thebibliography}{30}
\bibitem{FN}
Note that previous numerical studies 
did not use biorthonormalized eigenvectors in the parallel transport.
The thus obtained geometric factors differ from the present ones without contradicting them.
\end{thebibliography}
\end{document}